\documentclass[prl,aps,showpacs,twocolumn]{revtex4-1}
\usepackage{dcolumn}
\usepackage{bm}
\usepackage{amsmath,amssymb,amsfonts,latexsym,fancyhdr,graphicx,epstopdf,times,txfonts}

\begin{document}
\title{Non-Markovian qubit dynamics induced by Coulomb crystals}
\author{Massimo Borrelli$^1$, Pinja Haikka$^{2}$, Gabriele De Chiara$^3$, and Sabrina Maniscalco$^{1,2}$}
\affiliation{$^1$ CM-DTC, School of Engineering \& Physical Sciences, Heriot-Watt University, Edinburgh EH14 4AS, United Kingdom\\
$^2$Turku Centre for Quantum Physics, Department of Physics and Astronomy, University of
Turku, FI-20014 Turun yliopisto, Finland\\
$^3$ Centre for Theoretical Atomic, Molecular and Optical Physics, School of Mathematics and Physics, QueenÕs University Belfast, Belfast BT7 1NN, United Kingdom}

\begin{abstract}
We investigate the back-flow of information in a system with a second-order structural phase transition, namely a quasi one-dimensional Coulomb crystal. Using standard Ramsey interferometry which couples a target ion (the system) to the rest of the chain (a phononic environment), we study the non-Markovian character of the resulting open system dynamics. We study two different time-scales and show that the back-flow of information pinpoints both the phase transition and different dynamical features of the chain as it approaches criticality. We also establish an exact link between the back-flow of information and the Ramsey fringe visibility.

\end{abstract}

\maketitle

{\itshape Introduction} -- It has recently been suggested that an open system approach \cite{breuer} could be useful when studying properties of condensed matter physics, especially for systems presenting critical features. In this approach the open system is regarded as a quantum probe for the interacting many-body environment. The idea, then, is to deduce some of the properties of the environment by monitoring the reduced dynamics of the probe only \cite{pinja1,pinja2,pinja3}. Here we apply this framework to Ramsey interferometry using a single spin in a Coulomb crystal \cite{gabriele2,gabriele3,morigi1,morigi2}. A Coulomb crystal can be depicted as a low-dimensional structure composed of several dozens of ions trapped in an anisotropic potentials with very large transverse confinement \cite{dubin}. This kind of atomic spatial arrangement is routinely achievable in a standard Paul and Penning trap \cite{birkl,waki,raizen,ghosh}. Interestingly, Coulomb crystals exhibit a $T=0$ structural phase transition from a linear to a zig-zag equilibrium configuration. Such a transition can be tuned by a single experimentally controllable parameter, such as ion density or aspect ratio of the trap \cite{gabriele1}. In this Article we present a method for accurately pinpointing the critical parameter determining the phase transition.

It has been shown that by time-monitoring the interferometric signal of the Ramsey protocol one can measure the autocorrelation function of the crystal and detect the phase transition \cite{gabriele2}. Although this scheme is conceptually simple and experimentally feasible, two interesting questions are left open. If we look at the Ramsey protocol from an open system perspective, how does the dynamic of the single spin look like? Is there a way to signal the phase transition which would still rely on dynamical probing but, at the same time, be time-independent? Ramsey interferometry is made up of different unitary steps, making it impossible to define a unique Hamiltonian describing the whole probe-chain interaction. Nevertheless, one can imagine the single spin time-evolution as carried out inside a 'black box' and wonder about specific features of the consequent dynamical process. In this respect, we believe that a key feature addressing these questions is the (non-)Markovian character of the Ramsey interferometric scheme.

Interaction between a single quantum system and its environment leads to decoherence and dissipation, and the understanding of such processes is of primary importance for both foundational and applicative reasons. The general theory of open quantum systems draws a line separating two major classes of processes: Markovian and non-Markovian. In a Markovian process the system irreversibly looses all its quantum properties to the environment. A rigorous characterization for this class exists, based on the ability of describing the reduced system dynamics in terms of a semi-group of CPT dynamical maps \cite{lind}. A general definition of non-Markovian dynamics is more subtle and, as such, it has been subject of intense recent studies \cite{elsi,plenio,lu}. Loosely speaking, non-Markovian behavior is characterized by memory effects, enabling the system to temporarily recover some of its genuine quantum properties. Here we focus on a novel definition for the non-Markovian character of an open system dynamics, originally introduced in Ref. \cite{elsi}. The key idea to associate the distinguishability of two evolving system states to information flowing between the system and the environment: decrease in the state distinguishability implies information flux from system to environment, while a reversed direction of information flow temporarily increases state distinguishability. This definition, and a subsequent measure quantifying the amount of non-Markovianity in a dynamical process, has been successfully applied to investigate memory effects in a number of different physical contexts, from photonic systems \cite{lieu,jian} to cold atoms \cite{pinja2, pinja3} and spin-chains \cite{pinja1,mauro,mauro2}. 

In this Article we investigate the dynamical map associated to Ramsey interferometry in a quasi-1D Coulomb crystal and study its non-Markovian behavior as the system approaches a phase transition. Our results show that the distinguishability-based measure of non-Markovianity is very sensitive to the critical point and, in this sense, can be used to probe the structural phase transition of the Coulomb crystal and pinpoint the associated critical parameter unambiguously. Moreover, we find a direct link between this non-Markovianity measure and the Ramsey fringe visibility, providing an experimentally feasible way for a novel characterization of this critical phenomenon.

{\itshape Coulomb crystal} -- A linear Coulomb chain is composed of $N$ ions with mass $m$ and charge $Q$ interacting via a repulsive Coulomb potential and confined inside a highly anisotropic trap. Let $\nu$ and $\nu_{t}$ be the axial ($x$ direction) and the transverse ($y-z$ plane) trap frequencies. As shown in Ref. \cite{gabriele1}, if we assume ring-type boundary conditions, a second-order structural phase transition form a linear to a planar zig-zag configuration takes place when 
$\nu_{t}<\omega_{0}\sqrt{\frac{7}{2}\zeta(3)}\equiv\nu_{c}$,
where $\omega_{0}=\sqrt{Q^{2}/ma^{3}}$ and $a$ is the equilibrium inter-ion distance.
For $\nu_{t}<\nu_{c}$ the ionic equilibrium positions are $\vec{r}_{i}=(ia,0,0)$, with $i=1,2,\dots,N$. The ions oscillate harmonically around $\vec{r}_{i}$ resulting in small axial and transverse displacements from the equilibrium positions. In this spatial arrangement, the vibrational degrees of freedom of the chain are naturally uncoupled from each other. The normal modes of the chain can be labeled by the momentum $k=2\pi n/Na$ with $n=0,1,\dots,N/2$ and by their parity. $\nu_{t}$ is the largest transverse frequency of the normal modes excitation spectra.
On the other side of the phase transition the ions reorganize in a zig-zag spatial configuration in the $x-y$ plane where the new equilibrium positions are $\vec{r}_{i}=(ia,(-1)^{i}b/2,0)$. The transverse displacement $b$ depends on both $a$ and $\nu_{t}$ as shown in Ref. \cite{gabriele1}. In this case the $x$ and $y$ vibrational degrees of freedom of the chain are coupled and the first Brillouin zone is now $[0,\pi/2a]$, with $k=2\pi n/Na$ and $k=0,1,\dots,N/4$. The excitation spectra, which displays multiple branches, are far more structured than the linear case. For a detailed derivation and description of the system see Ref. \cite{gabriele1}. \\
{\itshape Dynamics of the probe} -- In the following we concentrate on one specific ion of the chain, labelled $j=1$ for convenience, and study its dynamics. Two internal states of this ion, $\{|e\rangle, |g\rangle\}$, form an open qubit system, and the rest of the ionic chain is taken to be its environment $E$. Since the relevant degrees of freedom of $E$ are the vibrational ones, we can use the normal modes decomposition and picture $E$ as a bosonic environment at $T=0$ with a non-trivial discretized excitation spectrum.
The qubit system $S$ couples to the environment $E$ via a resonant laser pulse that engineers the following interaction
\begin{equation}
\hat{H}_{\textrm{INT}}=\hbar\Omega\left[\hat{\sigma}^{+}e^{-i(\omega_{L}t-k_{L}\hat{y}_{1})}+\textrm{h.c.}\right],
\label{hint}
\end{equation}
where $\hat{\sigma}^{+}=|e\rangle\langle g|, \hat{\sigma}^{-}=|g\rangle\langle e|$, $\Omega$ is the Rabi frequency of the laser, $\omega_{L}, k_{L}$ are the laser frequency and wave-vector respectively,
and $\hat{y}_{1}$ is the ion position operator. The general  expression for $\hat{y}_{1}$ can be found in Ref. \cite{gabriele2}.
The target ion receives a state-dependent mechanical 'kick' in the $y$ direction, starts oscillating and excites
all the transverse vibrational normal modes of the chain. 
Operator $e^{-ik_{L}\hat{y}_{1}}=\bigotimes_{j}\exp\left(\alpha_{j}\hat{b}^{\dagger}_{j}-\alpha^{*}_{j}\hat{b}_{j}\right)$ is a displacement operator for the transverse normal modes of the chain, where $\hat{b}^{\dagger}_{j}, \hat{b}_{j}$ are the creation and annihilation operators for the $j-$th
normal mode of the chain, $\alpha_{j}=i\eta\sqrt{\omega_{0}/2\omega_{j}}S_{1j}$, Lamb-Dicke parameter is $\eta=k_{L}\sqrt{\hbar/m\omega_{0}}$ and matrix $S_{ij}$, which realizes the normal modes decomposition, is defined in Ref. \cite{gabriele2}. The $j$ index encodes all the quantum numbers defining the modes of the environment (momentum and parity).
The Ramsey interferometric protocol we implement goes as follows \cite{gabriele2}.
Assuming $\omega_{M} T_{L}\ll1$, where $\omega_{M}$ is the largest eigenfrequency of the total system and $T_{L}$ the laser pulse duration,
we set $\Omega T_{L}=\pi/4$ in Eq.\eqref{hint}, therefore applying a $\pi/2$ pulse, after which we let $S$ and $E$ evolve freely for a time $t$. The time-dependent dynamics is then governed by the following Hamiltonian
\begin{equation}
\hat{H}_{0}=\frac{\hbar\hat{\sigma}_{z}}{2}+\sum_{j}\hbar\omega_{j}\hat{b}^{\dagger}_{j}\hat{b}_{j},
\label{h0}
\end{equation}
where $\omega_{j}$ is the transverse frequency of the $j-$th mode.
Finally, we apply a $-\pi/2$ pulse.
It is important to remark that $t$ is the time elapsed between the two pluses.
Using the Bloch-sphere representation $(\theta,\phi)$ for $S$, we choose a generic pure initial state of the form
\begin{equation}
|\psi_{i}\rangle=\left[\cos\left(\frac{\theta}{2}\right)|e\rangle+e^{i\phi}\sin\left(\frac{\theta}{2}\right)|g\rangle\right]\bigotimes|\underline{0}\rangle,
\label{is}
\end{equation}
where $|\underline{0}\rangle$ is the total phononic vacuum state, which is defined as $|\underline{0}\rangle=\bigotimes_{j}|0_{j}\rangle$.
The final system-environment state reads
\begin{equation}
|\psi_{f}\rangle=\frac{1}{2}\left[|g,\chi_{g}(t)\rangle+|e,\chi_{e}(t)\rangle\right],
\label{fs}
\end{equation}
where $|\chi_{g}(t)\rangle, |\chi_{e}(t)\rangle$ are conditional environment states.
Since we are interested in the reduced system dynamics only we trace out the environment degrees of freedom to get the state of the probe qubit $\hat{\rho}_{S}(r)=\rho_{ee}(t)|e\rangle\langle e|+\rho_{gg}(t)|g\rangle\langle g|+\rho_{eg}(t)|e\rangle\langle g|+\textrm{h.c.}$

The overall process admits two complementary viewpoints: If we set the initial time $t_{0}=-T_{L}$, then
the system time-evolution is dictated by a complex quantum map, to which we cannot assign a unique global Hamiltonian describing its action. If, instead, we set $t_{0}=0$ we focus on the free evolution of an initially correlated system-environment state. Either way, we are effectively engineering and simulating a non-trivial open system dynamics where both dissipation and decoherence processes can take place. In the following we will quantify the degree of non-Markovinaity associated to this dynamical process.

{\itshape Non-Markovianity and criticality} -- We start by recalling the non-Markovianity measure $\mathcal{N}$ introduced in Ref. \cite{elsi}. Given two quantum states $\hat{\rho}_{1}, \hat{\rho}_{2}$
the distinguishability between them is quantified by the trace-distance $D=\textrm{tr}\left|\hat{\rho}_{1}-\hat{\rho}_{2}\right|/2$.
A generic Markovian dynamical map $\hat{\rho}(t)=\Lambda_{t}\hat{\rho}(t_{0})$ is contractive, that is, it continuously decreases the trance distance between two initially distinguishable quantum states. 
If at least two initial states exist such that this condition is violated for some time intervals then the map is non-Markovian. 
The non-Markovianity of a dynamical process described by a map $\Lambda_{t}$ can be quantified by the following integral
\begin{equation}
\mathcal{N}(\Lambda)=\max_{\hat{\rho}_{1,2}(0)}\int_{\sigma>0}dt\sigma(t,\hat{\rho}_{1,2}(0)),
\label{nm}
\end{equation}
where $\sigma(t,\hat{\rho}_{1,2}(0))=\frac{d}{dt}D(\hat{\rho}_{1}(t),\hat{\rho}_{2}(t))$ is the rate of change of the trace distance, that is, the information flux.
As shown in Ref. \cite{antti}, the maximization in Eq.\eqref{nm} can be restricted to pairs of initially pure orthogonal states $\hat{\rho}_{1}(0)$, $\hat{\rho}_{2}(0)$ for the case of a two-level system. These states, which lie on the boundary of the state space, are represented by antipodal points of the Bloch-sphere surface. This also explains the ansatz \eqref{is} for the initial state of the probe.
We now study the behavior of $\mathcal{N}$ as a function of the tuning parameter $\Delta=\nu_{t}/\nu_{c}-1$.
When the number of ions is large an exact maximization in Eq.\eqref{nm} is unfeasible. However, we have shown numerically that the maximizing pair is formed by the eigenstates of $\sigma_{x}$, which we label $|+\rangle, |-\rangle$. The corresponding trace distance reads as $D_{opt}(t)=\big|\rho_{eg}^{(+)}(t)-\rho_{eg}^{(-)}(t)\big|$. Interestingly, the dynamics of $|+\rangle, |-\rangle$ is purely dephasing, i.e., the optimal pair does not exchange excitations with the environment. 
In the following we show that the dynamics of the trace distance displays a sensitivity to the tuning parameter $\Delta$: the closer to criticality the chain is, the more damped the time-oscillations of $D$ are. We have shown that the trace distance is a function of the Ramsey fringe visibility $\mathcal{V}(t)$ \cite{gabriele2}.: 
\begin{equation}
D_{opt}(t)=\frac{1}{4}\left|1+2\cos\left[B(t)\right]\left(\mathcal{V}(t)-\frac{\xi^{4}}{\mathcal{V}(t)}\right)+\mathcal{V}^{4}(t)+2\xi^{4}\right|,
\label{tdinv}
\end{equation}
where $\xi=e^{-\sum_{j}|\alpha_{j}|^{2}/2}$ and $B(t)=\sum_{j}|\alpha_{j}|^{2}\sin(\omega_{j}t)$. This connection is one of our main results, establishing an exact analytical link between the experimentally measurable fringe visibility $\mathcal{V}(t)$ and the non-Markovianity $\mathcal{N}$ of the probe qubit. 
Let us remark that the above formula is general and specific of the Ramsey protocol only. The nature of the chain, that is either linear or zig-zag, is fully encoded in $\xi$ and $B(t)$.\\
\begin{figure}
\begin{center}
\includegraphics[width=0.4\textwidth]{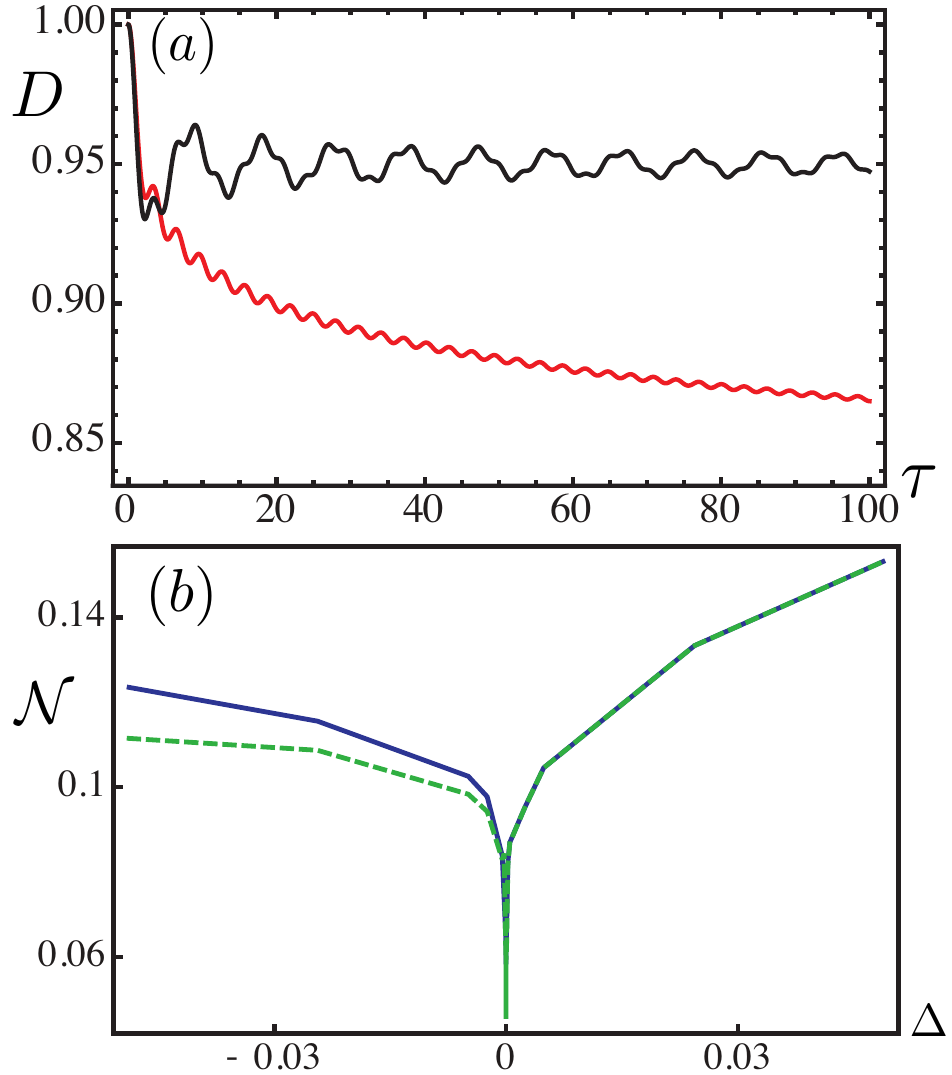}
\end{center}
\caption{(Color online) Short-time-scale dynamics.
(a) Time-evolution of the trace distance for the maximizing pair $\{|+\rangle, |-\rangle\}$ for $N=200$, far from the critical point (black line, $\Delta=0.1$) and very close to it (red line, $\Delta=10^{-5}$). (b) Non-Markovianity measure $\mathcal{N}$ for short time-scale truncation as a function of $\Delta$ for $N=100$ (blue solid line) and $N=1000$ (green dashed line).}
\label{plot1}
\end{figure}
{\itshape Short time-scale} -- We first analyze the short time behavior of the trace distance. In an experiment-based spirit, we set an upper bound $t_{M}$ to the elapsed time, which roughly corresponds to $250\mu$s (this time will also be the upper integration limit in \eqref{nm}) \cite{birkl}, define $\tau=\omega_{0}t$ and look at the time-evolution of the trace distance of the maximizing pair, Fig.\ref{plot1}(a). 
The behavior of $\mathcal{N}$ as a function of $\Delta$ is shown in Fig.\ref{plot1}(b) for $N=10^{2}, 10^{3}$. Coordinate $\Delta=0$ represents the critical point and the maximum distance from it that we consider is $10^{-1}$, whereas the minimum is $10^{-7}$.
There is a clear and rather abrupt change in the behavior of $\mathcal{N}$ at the critical point, which coincides with the appearance of the structural phase transition. Values of $\mathcal{N}$ are different on different sides of the critical point. However, approaching criticality both $\mathcal{N}_{\Delta\to0^{-}}$ and $\mathcal{N}_{\Delta\to0^{+}}$ converge to the same value, a non-zero absolute minimum.  This result fully reflects the time-evolution of the trace distance, as shown in Fig.\ref{plot1}(a). As the environment approaches the critical point the dynamics of the probe  becomes more and more damped, resulting in a Markovian behavior. Interestingly, the behavior shown in Fig.\ref{plot1}(b) is characteristic of a second order phase transition. Indeed, even if $\mathcal{N}$ is a continuous function when $\Delta\to0^{\pm}$, its derivative is clearly not. A similar feature is found when one studies derivatives of thermodynamical and statistical quantities in presence of a classical and quantum phase transition. We conclude this section by mentioning that the contribution of anharmonic terms in this regime does not greatly affect the above results. Discrepancies between the harmonic model of Eq.\eqref{h0} and a Landau-theory-based expansion of the interaction potential up to the $4^{\textrm{th}}$ order, are of the order of $10^{-3}$ for $\Delta\approx10^{-2}$.
\begin{figure}
\begin{center}
\includegraphics[width=0.4\textwidth]{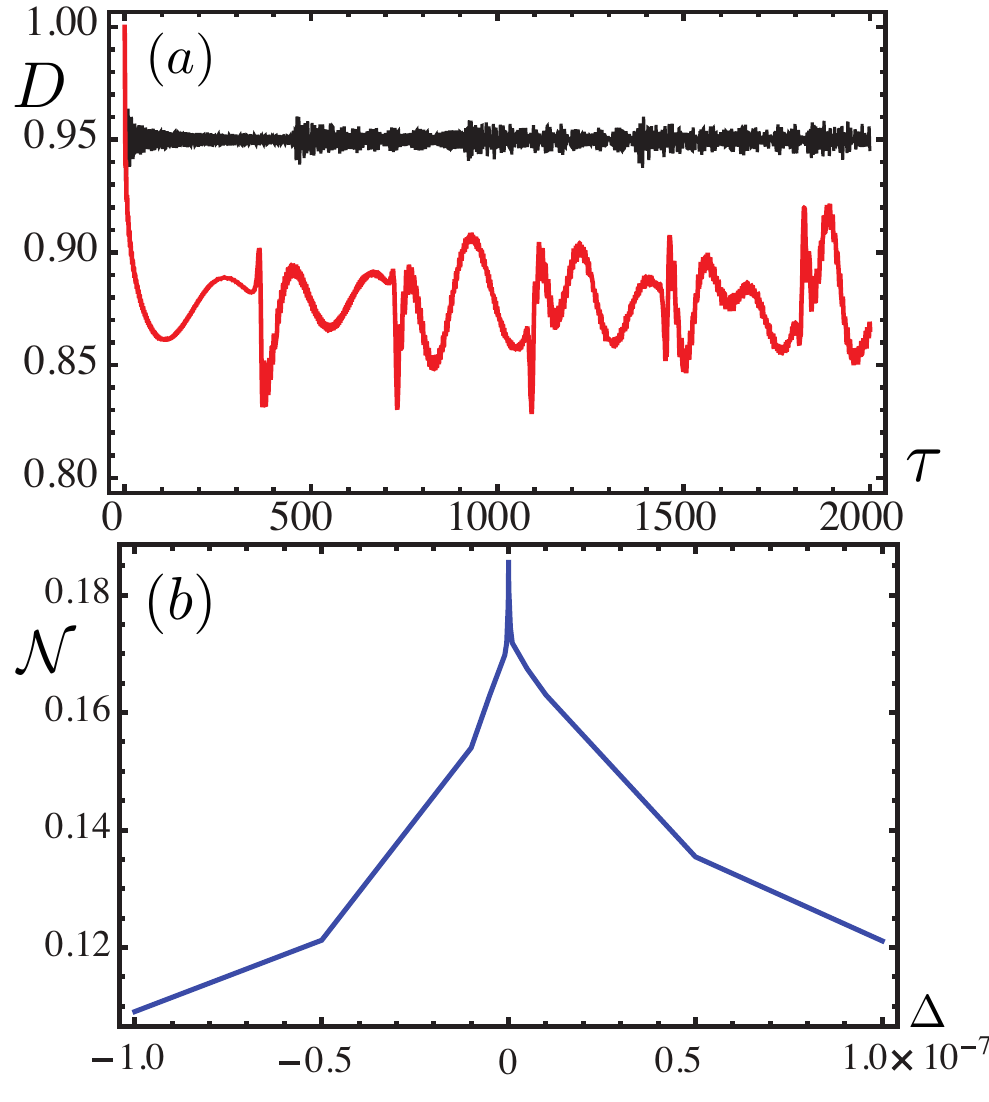}
\end{center}
\caption{(Color online) Long-time-scale dynamics. (a): time-evolution of the trace distance for the maximizing pair $\{|+\rangle, |-\rangle\}$, for $N=300$, far from the critical point (black, $\Delta=0.1$) and very close to it (red,  $\Delta=10^{-4}$). (b): Non-Markovianity measure $\mathcal{N}$ for long time-scale truncation as a function of $\Delta$ for $N=300$.}
\label{plot2}
\end{figure}

{\itshape Long time-scale in the thermodynamic limit} -- The next step is the investigation of the long-time scale behavior, corresponding to $t_{M}\approx5$ ms. In this regime, when the environment is pushed closer to criticality, anharmonic terms, arising from a Ginzburg-Landau expansion of the chain Hamiltonian become relevant \cite{gabriele1,gabriele4}. In this time regime, the full ab-initio calculation of the collective dynamics of a Coulomb chain is still an open problem. Since we are primarily interested in the qualitative behavior of \eqref{nm} close to criticality, we neglect anharmonic contributions. Within the range of validity of Ginzburg-Landau theory, it was shown in \cite{gabriele1,gabriele4} that the $4^{\textrm{th}}$ order amplitude $V^{(4)}$ close to criticality scales as $V^{(4)}\approx \omega_{0}^{2}/Na^{2}$, whereas $\omega\approx\omega_{0} \Delta+\delta k$ for all the relevant modes $q=\pi/a-\delta k$ near the soft mode. The thermodynamic limit is defined via the condition that, for $N\to\infty$, $a$ remains constant. Close to criticality, $\omega$ can become small but still finite and, depending on $N$, greater than $V^{(4)}$. The following results are intended to provide a qualitative picture of the long-time-scale regime in the thermodynamic limit.
The dynamics of the trace distance in this case is displayed in Fig. \ref{plot2}(a). Contrary to what happens in the short-time-scale regime, the closer $\Delta$ to zero, the larger the amplitude of the slowly damped revival peaks in $D$. Consequently, we expect the non-Markovianity measure to increase as the chain approach the critical point. Fig.\ref{plot2} fully confirms this prediction as $\mathcal{N}$ displays a cusp-like maximum when $\Delta\to0^{\pm}$. 
Close to criticality the mechanical instability of the chain is driven by the so-called soft mode \cite{gabriele1}. This mode is associated to the smallest frequency possible in the environment spectrum, which corresponds to $k=\pi/a$ and vanishes as $\Delta\to0^{\pm}$. 
Hence, close to the phase transition and for long enough times, the coupling between this almost stationary mode and the single spin will dominate over all the others, leading to a weakly damped oscillatory dynamics of the probe.
The action of the phononic background, which would force the system to lose information at short-time-scales as a consequence of the instability, is here overruled by an effective one-to-one coupling  between the target ion and the soft mode.
It is important to remark that the short and long time-scale regimes are totally compatible, with the common feature being the strong coupling between the probe and the soft mode at criticality. The main difference lies in the fact that for short times the probe is able to resolve only the high-energy part of the environment spectrum. 
On the contrary, if we wait sufficiently long time the revivals in the dynamics of the trace distance, due to the coupling with the soft mode, will appear leading to a the peak in non-Markovianity. Again, we stress that this result is a qualitative picture of the thermodynamic regime.
\begin{figure}[t!]
\begin{center}
\includegraphics[width=0.4\textwidth]{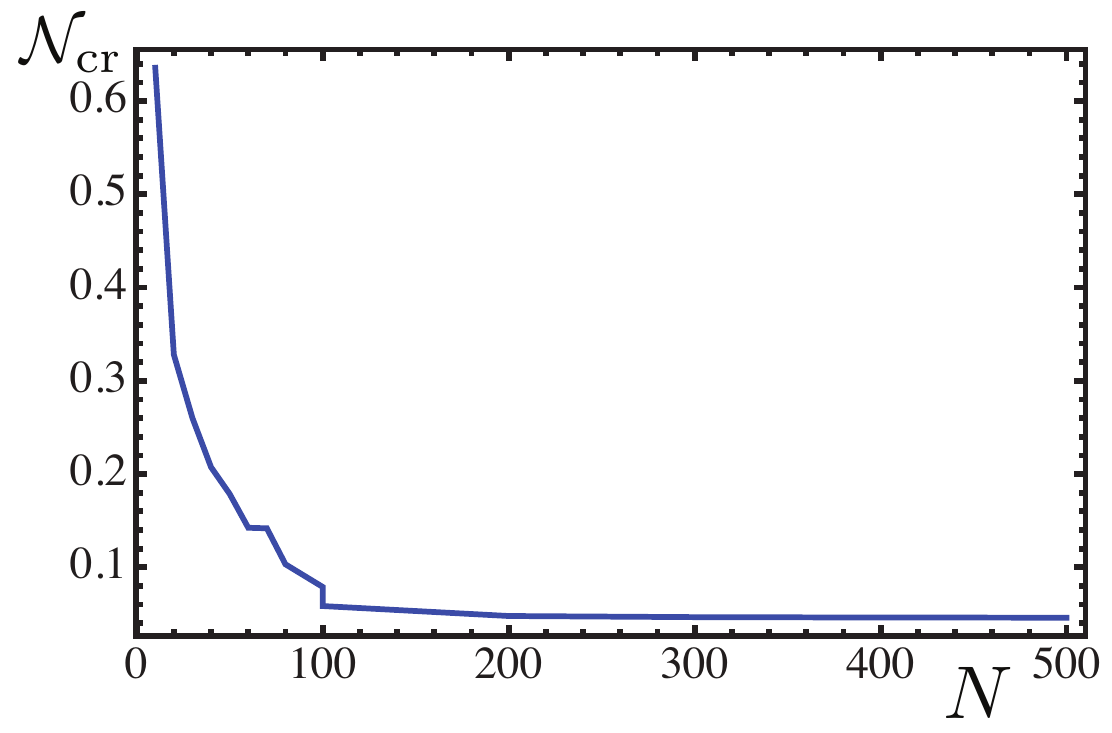}
\end{center}
\caption{(Color online) Short time-scale non-Markovianity measure $\mathcal{N}_{cr}$ at criticality as a function of the number of ions $N$.}
\label{plot3}
\end{figure}

{\itshape Finite size effects} -- Finally we address how finite size effects affect our results. To this aim we study how the short-time scale non-Markovianity measure scales with the number of ions in the chain close to critical point. The results are reported in Fig. \ref{plot3}. $\mathcal{N}_{cr}$ saturates to a non-zero value very soon. Finite-size effects, which lead to a larger value of the non-Markovianity measure, are relevant only for relatively small $N$ and no appreciable variation of $\mathcal{N}_{cr}$ is detectable for $N>100$. This is an indication that even in thermodynamical limit, the flow of information from the probe qubit to the rest of the chain is not complete. Loosely speaking, a minimum amount of the information initially stored in the probe remains trapped and does not get lost ever. As previously pointed out, for short times only the high-energy part of the spectrum (small $k$) contributes to the dynamics of the probe. This portion of the environment spectrum is essentially flat ($\partial\omega(k)/\partial k|_{k=0}=0$) and so all these modes dephase and rephase almost in sync; the system only leaks information. However, this happens on average only as the dephasing and rephasing cycles repeat many times within the short time interval we consider. This, in turn, implies that some of this information cannot flow out. It could do so completely if there were not the short-time-scale recurrences. These are associated to mechanical excitations going back to the system, resulting in revival peaks in the dynamics of the trace distance. The typical time for these revivals was estimated as $\tau\approx140$.

{\itshape Conclusions} -- We have shown that the back-flow of information in single-spin Ramsey interferometry is sensitive to the linear-to-zig-zag structural phase transition of a quasi 1-D Coulomb crystal. Abrupt changes in the behavior of $\mathcal{N}$ are observed when $\nu_{t}\to\pm\nu_{c}$, leading to extrema which clearly pin-point criticality. The nature of these extrema consistently reflects the dynamical features of the chain, which are responsible for driving the phase transition. We stress that the above analysis could be experimentally feasible. By time-monitoring the visibility of the Ramsey fringes, one can compute the non-Markovianity $\mathcal{N}$ of the interferometric scheme using Eq.\eqref{tdinv}. This method relies on dynamical probing and, at the same time, characterizes the phase transition in a time-independent way. 
\\
\\
We acknowledge financial support from EPSRC (EP/J016349/1), the Finnish Cultural Foundation (Science Workshop on Entanglement), the Emil Aaltonen Foundation (Non-Markovian quantum information) and the Magnus Ehrnrooth Foundation. M.B. would like to thank EPSRC CM-DTC for financial support, E.-M. Laine and M. Palma for useful discussions.

\end{document}